\documentclass[aps,twocolumn,showpacs]{revtex4-1}
\usepackage{graphicx}
\usepackage{amsmath}
\usepackage{amsfonts}
\usepackage[table]{xcolor}
\usepackage[caption=false]{subfig}
\usepackage{array}
\usepackage{floatrow}
\usepackage{pbox}
\usepackage{chngcntr}
\usepackage[colorlinks=true,linkcolor=blue,citecolor=blue,filecolor=cyan,urlcolor=blue,breaklinks=true]{hyperref}

\renewcommand{\vec}[1]{\boldsymbol{#1}}
\newcommand{\mean}[1]{\left\langle #1 \right\rangle}

\newcommand{\Fe}{\vec{F}^\mathrm{ext}}

\newcommand{\diag}{\operatorname{diag}}
\newcommand{\tr}{\operatorname{tr}}
\newcommand{\dbar}{d\hspace*{-0.08em}\bar{}\hspace*{0.1em}}

\begin{document}
\title{Universal trade-off between power, efficiency, and constancy\\ in steady-state heat engines}

\author{Patrick Pietzonka and Udo Seifert}
\affiliation{ II. Institut f\"ur Theoretische Physik, Universit\"at Stuttgart,
  70550 Stuttgart, Germany}
\date{\today}

\parskip 1mm

\begin{abstract}
  Heat engines should ideally have large power output, operate close to Carnot
  efficiency and show constancy, i.e., exhibit only small fluctuations in this
  output.  For steady-state heat engines, driven by a constant
  temperature difference between the two heat baths, we prove that out of
  these three requirements only two are compatible. Constancy enters
  quantitatively the conventional trade-off between power and
  efficiency. Thus, we rationalize and unify recent suggestions for overcoming
  this simple trade-off. Our universal bound is illustrated for a
  paradigmatic model of a quantum dot solar cell and for a Brownian gyrator
  delivering mechanical work against an external force.
\end{abstract}


\maketitle

The efficiency $\eta$ of heat engines operating cyclically between a hot heat reservoir at temperature $T_{h}$
and a cold one at $T_{c}$ is universally bounded from above by the Carnot
value $\eta_\mathrm{C}\equiv 1-T_{c}/T_{h}$. Moreover, it was commonly believed that reaching Carnot
efficiency inevitably comes with a vanishing power $P$ since such engines require
quasi-static conditions leading to an infinite cycle-time. 
This supposedly universal trade-off between power and efficiency has more recently been
challenged
by studies hinting at the possibility to come at least arbitrarily close to
Carnot efficiency with finite power by a particular coupling between subsystems \cite{alla13}
or by exploiting a working substance at a critical point \cite{camp16}. On the other hand, 
a general bound of the type $P\leq A (\eta_\mathrm{C}-\eta)$ with a system-specific amplitude $A$
has been proven for cyclic engines both in linear response \cite{bran15b} and beyond \cite{shir16}.
These two strands of reasoning can be reconciled only if one allows for a
diverging amplitude $A$ as the efficiency approaches the  Carnot value.

For steady-state heat
engines permanently coupled to two heat baths like in thermoelectric setups,
the common argument from above invoking quasistatic conditions and hence an infinite cycle time
is not directly applicable. It has usually been replaced with the idea that finite currents 
necessarily lead to dissipation which should spoil the option of reaching the Carnot limit
at finite power. This view was challenged by a seminal paper from Benenti \textit{et al.}\ \cite{bene11},
who pointed out that if time-reversal symmetry is broken, like in the presence of
a magnetic field, 
the usual approach based on linear irreversible thermodynamics does not forbid
Carnot efficiency at finite power. Subsequent studies showed with somewhat
more specific assumptions that Carnot efficiency at
finite power is not accessible. This holds true, \textit{e.g.}, for any finite 
number of terminals in a multiterminal set-up within linear response theory \cite{bran13a,bran15},
and for an effective two-terminal device containing inelastic electron-phonon processes \cite{yama16}.
However, as for cyclic engines, it has recently been pointed out that
in certain limits Carnot efficiency at finite power can be reached in models of
steady-state engines. Specific proposals
include two-cycle engines with diverging affinities \cite{pole17}, a specially
designed Feynman-Smoluchowski ratchet \cite{lee17}, and systems with a singular
transport law \cite{shir15a,koni16}.

These observations taken together indicate that when searching for a universal trade-off it
may not be enough to focus only on the \textit{two} characteristics
power and efficiency, and their relationship if one of them becomes maximal as it is typically
done, see also
\cite{curz75,vdb05,schm08,tu08,espo09,nakp10,seif11a,izum11,golu12a,toma13,star14,whit14,jian14,raz16,baue16,proe16}.
 As we will show in this Letter, as a crucial \textit{third} quantity,
 power fluctuations enter quantitatively into this balance.  Specifically, we will
prove for a huge class of steady-state heat engines, which includes all
thermodynamically consistent machines with a classical discrete set of
internal states and continuous ones modeled with an overdamped Langevin dynamics, that there is a universal trade-off
between three desiderata: Finite (or even large) power, an efficiency
close to the Carnot value, and constancy
in the sense of small fluctuations in the power output are not compatible.
Specifically, the bound
\begin{equation}
P\frac{\eta}{\eta_\mathrm{C}-\eta}\frac{T_{c}}{\varDelta_P} \leq \frac{1}{2}
\label{eq:b1}
\end{equation}
constrains (mean) output power, efficiency, and the power fluctuations
in finite time as measured by
\begin{equation}
\varDelta_P \equiv \lim _{t\to \infty} \langle (P(t)-P)^2\rangle t .
\end{equation}  Here, $P(t)$ is
the fluctuating power after time $t$ evaluated in the steady state,
for which $\langle ...\rangle$ denotes averages. Since the mean output work grows linearly
in time as does its variance, converting work fluctuations into power fluctuations
requires the additional factor of $t$ to reach a finite limit for $\varDelta_P$,
with which we  characterize the
constancy, or stability \cite{holu14},  of the engine. In particular
for nanoscopic heat engines, power fluctuations due to thermal noise are not
negligible compared to the
mean work output on relevant time scales and should therefore be taken into account for a
thermodynamic description.
The crucial role  power fluctuations play in the above bound is complementary
to their effect in the statistics of efficiency in a finite time
\cite{verl14,ging14,pole15,proe15}.

As a main first consequence of this new bound for steady-state heat engines,
it is obvious that as long as the power fluctuations remain finite (and
$T_{c}$ as well), approaching Carnot efficiency implies that the power has to
vanish at least linearly.  The explicit occurrence of $\varDelta_P$ as an
amplitude in a putative linear relationship between power and efficiency gap
to Carnot, however, offers a second option. If the fluctuations blow up, then
a finite (or even diverging) power is not ruled out as $\eta\to
\eta_\mathrm{C}$ which unifies quantitatively the various observations
recalled above.

The bound \eqref{eq:b1} can be rearranged as a bound on efficiency
\begin{equation}
\eta\leq\frac{\eta_\mathrm{C}}{1+2PT_{c}/\varDelta_P} 
\label{eq:b2}
\end{equation}
 determined by mean and 
fluctuations of the output power. Thus, the efficiency of any steady-state heat engine
is bounded from above by this simple expression independent of the specific design of
the engine.
The formal similarity of \eqref{eq:b2} with a bound derived for the efficiency of molecular
motors \cite{piet16b} indicates as common origin of these bounds  the thermodynamic
uncertainty relation \cite{bara15,ging16}, which describes a universal inequality between entropy production 
and mean and variance of an arbitrary current.

For a proof of the universal trade-off \eqref{eq:b1}, we consider an engine characterized by a set of internal states
$\{i\}$ with internal energies $\{E_i\}$. 
A transition between  state $i$ and $j$ takes place with a rate  $k^c_{ij}$ if it is
mediated by the contact to the cold bath and with a rate $k^h_{ij}$ if
it is mediated by the hot bath.
One of these rates can be zero which means that this particular transition always involves 
the other bath.
For any nonzero rate, the corresponding backward rate does not vanish
either, and it obeys 
the local detailed balance condition \cite{seif12}
\begin{equation}
\frac{k^{c,h}_{ij}}{k^{c,h}_{ji}}=\exp[(E_i-E_j+b_{ij}^{c,h}\mu_{c,h}-fd_{ij})/T_{c,h}],
\label{eq:k}
\end{equation}
where we set Boltzmann's constant to 1 throughout the Letter.
Here, $E_{i,j}$ are the internal energies of the two states. If the
the transition from
$i$ to $j$ requires transport of an electron
from  the bath with temperature $T_a$ $(a=c,h)$ and chemical potential $\mu_a$  to the system,
then  $b_{ij}^a=1$. 
Likewise, if this transition involves the release of an electron to a bath, $b_{ij}^a=-1$.
In both cases, the chemical potentials enter the
expression in the exponent providing a contribution to the total
free energy involved in  
such a transition. The last term in the exponent is non-zero if this transition
additionally involves a step
of length $d_{ij}$ against an external force $f$. Generalizations to several species of
particles, further baths, or the case of rotary motion against an applied torque
should be obvious.

In the steady state, where state $i$ is realized with probability $p_i$,
 this engine can
be characterized by three important mean currents, the heat current 
\begin{equation}
j_h= \sum_{i<j}(p_ik^h_{ij}-p_jk^h_{ji})(E_j-E_i-b^h_{ij}\mu_h)
\end{equation} from the hot bath to
the engine, the heat current 
\begin{equation}
j_c= -\sum_{i<j}(p_ik^c_{ij}-p_jk^c_{ji})(E_j-E_i-b^c_{ij}\mu_c)
\end{equation}
from the engine to the cold bath and the ``work current''
\begin{equation}
j_w=j_h-j_c=P, 
\label{eq:1stlaw}
\end{equation} which, due to the first law, is the power delivered by the engine.
Running the engine for a finite time $t$, each of the currents will 
fluctuate around these mean values with a dispersion \footnote{The fluctuating
currents $j_\alpha(t)$ are formally defined as
$j_h(t)=\sum_{i<j}n_{ij}^h(t)(E_j-E_i-b_{ij}^h\mu_h)/t$,
$j_c(t)=-\sum_{i<j}n_{ij}^c(t)(E_j-E_i-b_{ij}^c\mu_c)/t$, and
$j_w(t)=\sum_{i<j}[-b_{ij}^hn_{ij}^h(t)(\mu_h-\mu_c)+(n_{ij}^h(t)+n_{ij}^c(t))f
d_{ij}]/t$, with $n_{ij}^a(t)$ being the
net number of transitions between $i$ and $j$ mediated by the hot ($a=h$) or
cold ($a=c$) bath up to the time $t$.}
\begin{equation}
D_\alpha\equiv \lim_{t\to\infty} \langle (j_\alpha(t)-j_\alpha)^2\rangle t/2
\end{equation} where $\alpha=h,c,w$. The mean entropy production rate becomes
\begin{equation}
\sigma = j_c/T_{c}-j_h/T_{h}= j_w(\eta_\mathrm{C}/\eta -1)/T_{c}.
\label{eq:sigma}
\end{equation}
Since this Markovian network is  thermodynamically consistent,
one can directly apply the thermodynamic uncertainty relation, which
reads for any of the three currents \cite{bara15,ging16}
\begin{equation}
\sigma D_\alpha\geq  j_\alpha^2 .
\label{eq:unc}
\end{equation} Evaluating this relation for the work current, $\alpha=w$, 
leads, with  $\varDelta_P = 2 D_w$, \eqref{eq:1stlaw} and \eqref{eq:sigma},  to the bound \eqref{eq:b1}.

Two related, but not identical,  forms of this bound can be derived similarly
by applying the thermodynamic uncertainty
relation to either the heat current from the hot bath or the one entering the cold bath.
Expressed as bound on power, they read explicitly
\begin{equation}
P\leq(\eta_\mathrm{C}-\eta)\eta D_h/T_{c}
\label{eq:b3}
\end{equation}
and
\begin{equation}
P\leq(\eta_\mathrm{C}-\eta)\eta D_c/[(1-\eta)^2 T_{c}],
\label{eq:b4}
\end{equation}
respectively. 
Obviously, in order to reach a finite value for the power as $\eta\to\eta_\mathrm{C}$, the fluctuations
in all three currents have to diverge at least $\sim1/(\eta_\mathrm{C}-\eta)$ since each of the three
bounds (\ref{eq:b1},\ref{eq:b3},\ref{eq:b4}) has to be respected.

\begin{figure}
  \centering
  \sidesubfloat[]{\includegraphics[width=0.5\textwidth]{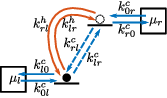}}
  \sidesubfloat[]{\includegraphics[width=0.3\textwidth]{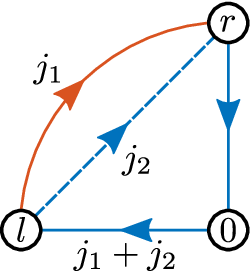}}\\%
  \sidesubfloat[]{\includegraphics[width=0.8\textwidth]{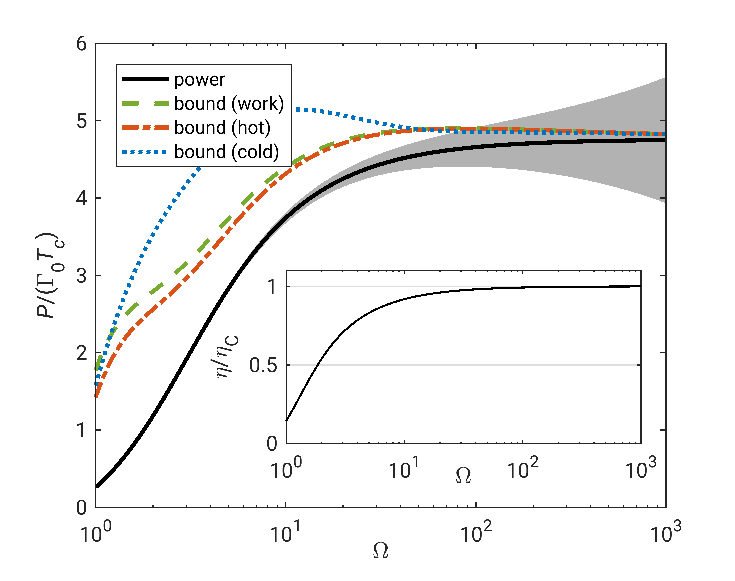}}
  \caption{(a) Model for a photoelectric device, transporting electrons
    between two reservoirs through a quantum dot with two energy
    levels. Transitions to and from the baths (blue solid arrows) and
    nonradiative transitions between the energy levels (blue dashed arrows)
    occur in contact with the cold bath. Radiative transitions (red
    curved arrows) occur in contact with the hot bath. (b) The network
    representation of the three possible states of the quantum dot allows for
    an identification of the two cycle currents associated with radiative
    ($j_1$) and nonradiative transitions ($j_2$). (c) Output power and
    efficiency (inset) of the photoelectric device as a function of the
    scaling parameter $\varOmega$
    entering the rates and affinities according to  $x_c=10$,
    $x_h=0.2$, $x_l=1$, $x_r=1.2+7/\varOmega$,
    $\varGamma_l=\varGamma_r=\varGamma_h=\varOmega\varGamma_0$,
    $\varGamma_c=\varOmega^{-1.5}\varGamma_0$, where $\varGamma_0$ is a frequency
    of reference. 
    The power fluctuations, quantified by
    $\sqrt{\varDelta_P/(10\varGamma_0)}$, are shown as a grey shaded region. The three bounds on the
    output power (\ref{eq:b1},\ref{eq:b3},\ref{eq:b4}) are shown as dashed,
    dash-dotted and dotted curves, respectively. }
  \label{fig:solarcell}
\end{figure}

For an illustration of these bounds, we consider a simple but instructive
model for a solar cell introduced in Ref.~\cite{rutt09} as shown in
Fig.~\ref{fig:solarcell}a.  It consists of a two level quantum dot that can
either be empty (state $0$) or contain an electron in one of the levels
$E_l<E_r$ (states $l$ and $r$, respectively).  Electrons are transported
through the dot from a left reservoir with chemical potential $\mu_l$ and
temperature $T_c$ to a right reservoir with higher chemical potential $\mu_r$
and the same temperature $T_c$.  They enter the level $E_l$ from the left
reservoir at a rate $k_{0l}^c$, and they jump back at a rate
$k_{l0}^c$. Analogously, the level $E_r$ is connected to the right reservoir
via the transition rates $k_{r0}^c$ and $k_{0r}^c$.  Transitions between the
two states can occur either nonradiatively through interactions with the
surrounding phonon bath at temperature $T_c$ with rates $k_{lr}^c$ and
$k_{rl}^c$ or are mediated by the black body radiation of the sun with
temperature $T_h$ at rates $k_{lr}^h$ and $k_{rl}^h$.  The transition rates to
and from the two electron reservoirs ($i=l,r$) are chosen according to the Fermi-Dirac
distribution as
\begin{equation}
  \label{eq:fd}
  k_{0i}^c=\varGamma_i/[1+\exp (x_i)],\ k_{i0}^c=\varGamma_i/[1+\exp (-x_i)]
\end{equation}
with frequencies $\varGamma_i$ and
affinities $x_i=(E_i-\mu_i)/T_c$. The transition rates between the two states
in contact with the cold phonon bath or hot photon bath ($a=c,h$,
respectively) are determined
by the Bose-Einstein distribution 
\begin{equation}
  \label{eq:be}
  k_{lr}^a=\varGamma_a/[\exp(x_a)-1],\   k_{rl}^a=\varGamma_a/[1-\exp(-x_a)], 
\end{equation}
with frequencies $\varGamma_j$ and
affinities $x_a=(E_r-E_l)/T_a$. All four pairs of rates satisfy the local
detailed balance condition~\eqref{eq:k}.

\floatsetup[figure]{style=plain,subcapbesideposition=top}

In a network representation of this system
(Fig.~\ref{fig:solarcell}b), we can identify two independent cycle
currents. The variables $X_1$ and $X_2$ count the net number of
electrons transported from state $l$ to state $r$ via radiative and
nonradiative transitions, respectively. The total number of
transported electrons is thus $X_{e}=X_1+X_2$ (up to a single electron
that might still be in the two-level system, which does not affect the
statistics of fluctuations in the long-time limit).  The joint
fluctuations of these counting variables are characterized by the
average currents $\mean{X_{1,2}(t)}=j_{1,2}t$ and their covariance
$\mean{(X_i(t)-j_it)(X_j(t)-j_jt)}=2D_{ij}t$ for large $t$, which can
be calculated directly from the rates \eqref{eq:fd} and
\eqref{eq:be}~\footnote{See Supplemental Material (including
  Refs. \cite{koza99,baie09c,touc17}) below.}.

For the mean power, mean heat current from the hot reservoir, and mean heat current into the
cold reservoir, one obtains
\begin{align}
  P&=\varDelta\mu\ (j_1+j_2),\label{eq:solarpower}\\
  j_h&=\varDelta E\ j_1,\\
  j_c&=(\varDelta E-\varDelta\mu)\ j_1-\varDelta\mu\, j_2,
\end{align}
respectively, 
with $\varDelta E\equiv E_r-E_l$ and $\varDelta \mu\equiv \mu_r-\mu_l$.
For the corresponding dispersion coefficients one gets
\begin{align}
  D_w&=\varDelta\mu^2(D_{11}+D_{22}+2D_{12}),\\
  D_h&=\varDelta E^2D_{11},\\
  D_c&=(\varDelta E-\varDelta\mu)^2D_{11}+\varDelta\mu^2D_{22}
-2(\varDelta E-\varDelta\mu)\varDelta\mu D_{12}.\label{eq:solardc}
\end{align}

In the ``strong coupling'' limit of negligible nonradiative transitions
($\varGamma_c\to 0$ and hence $X_2, j_2, D_{22}$, and $D_{12}\to 0$), the
efficiency of the photoelectric device becomes $\eta=\varDelta\mu/\varDelta
E$. By gradually increasing the chemical potential difference $\varDelta \mu$
with fixed $\varDelta E$, this ratio can approach $\eta_\mathrm{C}$ from
below. In Fig.~\ref{fig:solarcell}c, this limit is realized by increasing the
chemical potential $\mu_r$ via a scaling parameter $\varOmega$ while keeping
$\mu_l$, the energies and the temperatures fixed, as detailed in the
caption. We show that finite power can be achieved in this limit by increasing
the rates $\varGamma_l$, $\varGamma_r$ and $\varGamma_h$ while reducing
$\varGamma_c$ \footnote{A similar scaling has been used to achieve Carnot
  efficiency at divergent power output in Ref.~\cite{pole17}.}. Crucially, as
predicted by relation \eqref{eq:b1}, the fluctuations of the output power
diverge in such a scenario.

The uncertainty relation \eqref{eq:unc}
becomes tight in the linear response limit for unicyclic networks \cite{bara15}. Hence, for
strong coupling and vanishing cycle affinity
\begin{equation}
  \ln\frac{k_{0l}^ck_{lr}^hk_{r0}^c}{k_{0r}^ck_{rl}^hk_{l0}^c}=x_r-x_l-x_h=(\eta_\mathrm{C}\varDelta E-\varDelta\mu)/T_c,
\end{equation}
as realized  for a large scaling parameter $\varOmega$ in Fig.~\ref{fig:solarcell}c,
all three bounds on the power output (\ref{eq:b1},\ref{eq:b3},\ref{eq:b4}) saturate. 
Beyond the linear response and strong-coupling limit, the bounds can become
weaker. As explored in the Supplemental Material \cite{Note2}, depending on the values of
the transition frequencies, each of the three
bounds can become the strongest one demonstrating that these are independent bounds. 

\begin{figure}
  \centering
  {\includegraphics[width=\textwidth]{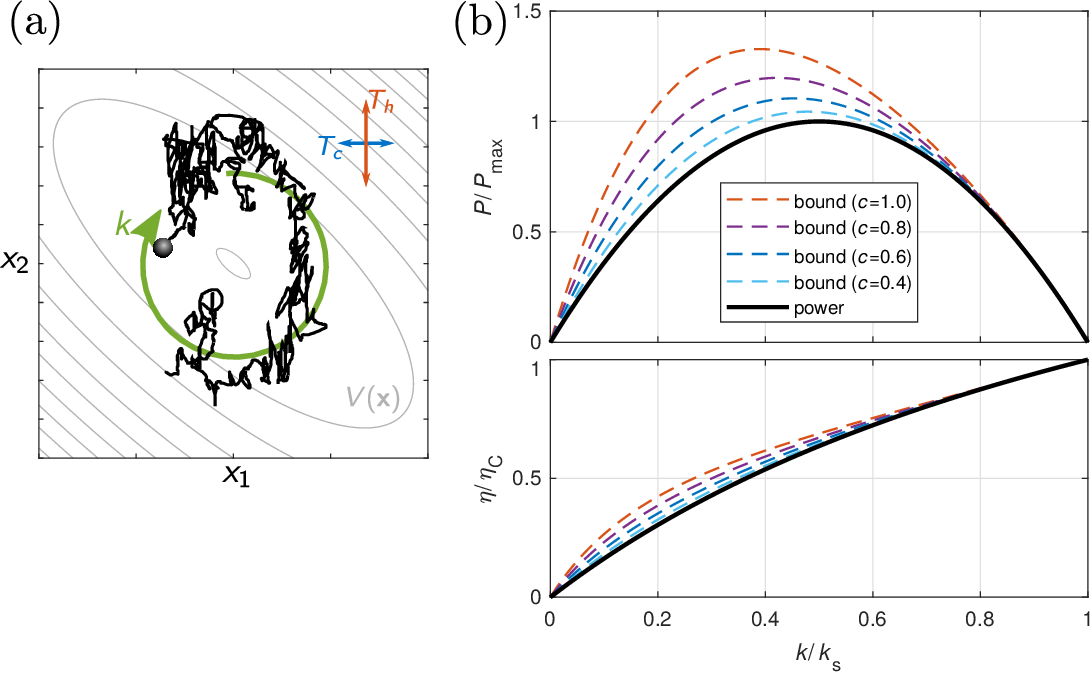}}
  \caption{(a) Schematic trajectory of a Brownian gyrator with the
    two-dimensional potential $V(x_1,x_2)$ shown as contour lines. The
    anisotropic noise leads to a gyration in counter-clockwise direction
    against the external force with parameter $k$. (b) Scaled power (top) and
    efficiency (bottom) as a function of the force parameter $k$. The
    respective bounds \eqref{eq:b3} and \eqref{eq:b2} become tighter with
    decreasing $c$.  Parameters: $T_c=1$, $T_h=7$, $\mu=1$, $u_1=1$,
    $u_2=1.2$.}
  \label{fig:gyrator}
\end{figure}
Since the thermodynamic uncertainty relation \eqref{eq:unc} has also been
proven for overdamped Langevin dynamics \cite{pole16,ging16a}, the resulting
trade-offs apply as well to steady-state heat engines modeled on such a
continuous state space. For an analytically solvable illustration, we consider
the so-called Brownian gyrator \cite{fill07}, which has recently been realized
experimentally in an electronic \cite{chia17} and in a colloidal
system \cite{argu17}. The setup, shown in Fig.~\ref{fig:gyrator}a, consists of
a point particle with mobility $\mu$ in two dimensions and an anisotropic
harmonic potential $V(\vec{x})=(u_1x_1^2+u_2x_2^2)/2+cx_1x_2$ with $u_{1,2}>0$
and $0<c<\sqrt{u_1u_2}$. Its motion obeys the Langevin equation
\begin{equation}
  \label{eq:langevin_gyrator}
  {d\vec{x}/dt}=\mu\left[-\partial
      V/\partial\vec{x}+\Fe(\vec{x})\right]+\vec{\zeta}(t),
\end{equation}
where the noise term has mean $\mean{\vec{\zeta}(t)}=0$ and correlations
$\mean{\zeta_i(t)\zeta_j(t')}=2\mu T_i\delta_{ij}\delta(t-t')$, i.e., the two
components of the fluctuating force are associated with heat reservoirs at two
different temperatures $T_i$. We choose $T_1\equiv T_c<T_2\equiv T_h$. In the
established case with $\Fe(\vec{x})=0$, the coupling $c\neq 0$ in the
potential mediates a steady transfer of heat from the hot to the cold reservoir,
that is accompanied by a persistent directed gyration of the particle \cite{fill07}. In
order to extract mechanical work from this gyration, we load the engine with
a nonconservative external counter force $\Fe(\vec{x})$. As the simplest, linear
realization of such a force, we choose $\Fe(\vec{x})=(F^\mathrm{ext}_1,F^\mathrm{ext}_2)^T=k\,(x_2,-x_1)^T$ with a
parameter $k$. Following the rules of stochastic thermodynamics \cite{seif12},
along an individual trajectory we identify the work
\begin{equation}
  \dbar w=-d\vec{x}\cdot \Fe(\vec{x})
\label{eq:dw}
\end{equation}
performed against the external force, the heat
\begin{equation}
  \dbar q_c=dx_1[-\partial V(\vec{x})/\partial x_1+F^\mathrm{ext}_1(\vec{x})]
\label{eq:dqc}
\end{equation}
dissipated into the cold heat bath, and the heat
\begin{equation}
  \dbar q_h=-dx_2[-\partial V(\vec{x})/\partial x_2+F^\mathrm{ext}_2(\vec{x})]
\label{eq:dqh}
\end{equation}
extracted from the hot heat bath. The corresponding integrated
currents follow from these differentials
through the Stratonovich integration scheme. 

The average mechanical power can be calculated as \cite{Note2} 
\begin{equation}
  \label{eq:gyrpower}
  P=j_w=\mean{\dbar w/dt}=4P_\mathrm{max}\frac{k}{k_s}\left(1-\frac{k}{k_s}\right)
\end{equation}
with the stall-parameter $k_s\equiv c\eta_C/(2-\eta_C)$ and the maximal
power $P_\mathrm{max}\equiv \mu k_s^2(T_c+T_h)/(2(u_1+u_2))$. The
system operates as a heat engine delivering mechanical work for
$0<k<k_s$.  Its efficiency is then given by
\begin{equation}
  \eta=j_w/j_h=2 k/(c+k)\leq\eta_C.
\label{eq:gyratoreff}
\end{equation}
The diffusion coefficients $D_{w,c,h}$ of the three currents can be calculated
analytically as well. It turns out that the
uncertainty relation \eqref{eq:unc} becomes the same for each current, 
\begin{equation}
  \label{eq:unc_gyr}
  \frac{D_w\sigma}{j_w^2}=  \frac{D_c\sigma}{j_c^2}=
  \frac{D_h\sigma}{j_h^2}=1+\frac{[c\eta_C(1-k/k_s)]^2}{(u_1+u_2)^2(1-\eta_C)}\geq 1,
\end{equation}
with the entropy production rate $\sigma$ as defined in
Eq.~\eqref{eq:sigma}.  The resulting bounds on the power and
efficiency are shown in Fig.~\ref{fig:gyrator}b. These bounds become
strong when the ratio $D_w\sigma/j_w^2$ in Eq.~\eqref{eq:unc_gyr} is
close to one, as it is the case for moderate temperature differences
or small couplings $c$ with respect to the parameters $u_{1,2}$. For
$k$ close to $k_s$, the bounds on both power and efficiency become
tight in linear order in $(k_s-k)$ independently of all other
parameters. Because of a tight coupling between heat and work
currents, this regime corresponds to a weak driving of the system
within the linear response regime, for which the uncertainty relation can
generally be saturated.  "For $k\to k_s$, as the efficiency approaches
the Carnot limit, the power vanishes. Since the constancy of the power
remains finite, this typical behavior of a heat engine can be
interpreted as a consequence of the universal
relation~\eqref{eq:b1}. The unusual case of finite power close to
Carnot efficiency can yet be obtained by scaling the mobility $\mu$,
which comes at the cost of diverging power fluctuations~\cite{Note2}.

In summary, we have derived a bound providing a universal trade-off between
power, efficiency and constancy, i.e., fluctuations in the power output. A
finite (or even diverging) power as Carnot efficiency is approached
necessarily requires the latter to diverge. The three versions of the bound
hold beyond the linear response regime of a small temperature difference
between the two heat reservoirs. We have derived these results for steady-state engines described by a thermodynamically consistent Markovian dynamics
on a discrete state space and for overdamped Langevin dynamics. They also
apply to apparently non-Markovian heat engines, provided that there is a
sufficiently fine-grained level of description on which the system obeys a
Markovian stochastic dynamics, however complex this description might be.  The
generalization to underdamped Langevin dynamics might be somewhat more subtle.

One might expect that similar results could be derived for cyclic, i.e.,
periodically driven heat engines \cite{schm08} that can be experimentally realized
for colloidal systems \cite{blic12,mart16} and with active baths \cite{kris16}.
In particular, a certain formal analogy of \eqref{eq:b1}
with the bound derived in \cite{shir16} for cyclic engines is striking. However, the analysis
 of a periodically driven Brownian clock in \cite{bara16} shows that
the steady-state bound \eqref{eq:unc} relating entropy production, mean current, and dispersion
cannot naively be extended to periodically driven systems. Therefore, it remains an
exciting open question whether there are  periodically driven heat engines that
beat the, for steady-state engines universal, bound \eqref{eq:b1}.

We have focused on the question whether Carnot efficiency can be
reached with finite power.  More generally, one can ask for any given
class of machines whether the power at the maximal efficiency, which
may be less than Carnot (or less than 1 for isothermal machines), can
be bounded or shown to vanish. Exploring this issue using the
techniques introduced here will be left to future work.  Finally, it
will be interesting to investigate whether and how constancy enters
bounds for genuine quantum heat engines that exploit coherences---see
\cite{bran17} and references therein.

\bibliography{/home/public/papers-softbio/bibtex/refs}

\clearpage

\onecolumngrid
\counterwithout{equation}{section}
\setcounter{section}{0}
\setcounter{equation}{0}
\renewcommand\theequation{S.\arabic{equation}}

\section*{Supplemental Material: Universal trade-off between power, efficiency,
  and constancy in steady-state heat engines}

\section{Quantum dot solar cell}
\label{sec:solarcell}

\begin{figure}[b]
  \centering
  \includegraphics[width=0.35\textwidth]{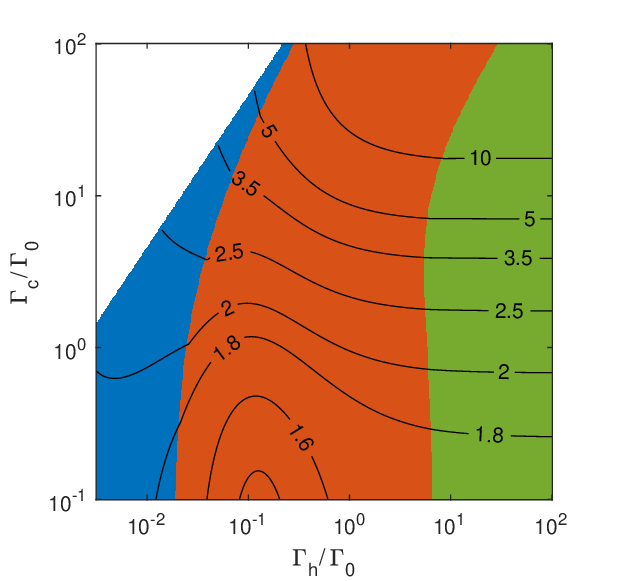}
  \caption{Colormap showing which of
    the three bounds on the output power is the tightest as a function of the
    transition frequencies $\varGamma_c$ and $\varGamma_h$, with an arbitrary
    frequency of reference $\varGamma_0$. In the white area
    in the top left corner the device does not operate as a heat engine. The
    colored areas from left to right refer to the bounds \eqref{eq:b4}
    (fluctuations of dissipated heat, blue), \eqref{eq:b3}
    (fluctuations of heat from the hot bath, red), and \eqref{eq:b1} (fluctuations of
    work, green). The contour lines show the ratio of the respective bounds to
    the actual power. Parameters: $x_l=0$, $x_r=3.8$, $x_h=0.1$, $x_c=7$,
    $\varGamma_l=\varGamma_0$, $\varGamma_r=10\varGamma_0$.}
  \label{fig:solarcell2}
\end{figure}

The joint fluctuations of the counting
variables $X_1$ and $X_2$  give rise to the scaled cumulant generating function
\begin{align}
  \lambda(z_1,z_2)&\equiv\lim_{t\to\infty}\frac{1}{t}\ln\mean{e^{z_1 X_1+z_2
    X_2}}\nonumber\\
&=j_1 z_1+j_2 z_2+D_{11}z_1^2+D_{22}z_2^2+2D_{12}z_1z_2+\mathcal{O}(z^3)
\label{eq:scgf}
\end{align}
with the mean currents $j_{1,2}=\mean{X_{1,2}}/t$ and the symmetric diffusion
matrix $D$. Technically, this function follows as the maximal eigenvalue of
the matrix 
\begin{equation}
  \mathcal{L}_{ij}(z_1,z_2)=k_{ji}^he^{d_{ji}z_1}+k_{ji}^ce^{d_{ji}z_2}-\delta_{ij}\sum_{\ell}(k_{i\ell}^c+k_{i\ell}^h)
\end{equation}
with $i,j\in\{0,l,r\}$, $d_{lr}=-d_{rl}=1$ and $d_{ij}=0$ otherwise.
The cumulants in the expansion \eqref{eq:scgf} can then be calculated using
the standard methods explained in Refs.~\cite{koza99,baie09c} as
\begin{align}
  j_1=[-k^c_{l0} k^c_{0r} k^h_{rl}+k^c_{0l} (k^c_{r0} k^h_{lr}+k^c_{rl}
   k^h_{lr}-k^c_{lr} k^h_{rl})+k^c_{rl} k^c_{0r} k^h_{lr}-k^c_{0r}
   k^c_{lr} k^h_{rl}]/\mathcal{N},
\end{align}
\begin{align}
j_2=[-k^c_{l0} k^c_{rl} k^c_{0r}+k^c_{r0} k^c_{0l}
   k^c_{lr}+(k^c_{0l}+k^c_{0r}) (k^c_{lr} k^h_{rl}-k^c_{rl}
   k^h_{lr})]/\mathcal{N},
\end{align}
\begin{align}
  D_{11}=&[-2 j_1^2
   (k^c_{l0}+k^c_{r0}+k^c_{0l}+k^c_{rl}+k^c_{0r}+k^c_{lr}+k^h_{rl}+k^h_{lr})+j_1 (2 k^c_{rl} k^h_{lr}-2 k^c_{lr} k^h_{rl})+k^c_{l0}
   k^c_{0r} k^h_{rl}+k^c_{r0} k^c_{0l} k^h_{lr}\nonumber\\
&+k^c_{0l} k^c_{rl}
   k^h_{lr}+k^c_{0l} k^c_{lr} k^h_{rl}+k^c_{rl} k^c_{0r}
   k^h_{lr}+k^c_{0r} k^c_{lr} k^h_{rl}]/(2\mathcal{N}),
\end{align}
\begin{align}
   D_{12}=&[-2 j_1 j_2
   (k^c_{l0}+k^c_{r0}+k^c_{0l}+k^c_{rl}+k^c_{0r}+k^c_{lr}+k^h_{rl}+k^h_{lr})+j_2(k^c_{rl} k^h_{lr}- k^c_{lr} k^h_{rl})\nonumber\\ 
   &+j_1(k^c_{lr} k^h_{rl}-k^c_{rl} k^h_{lr})-(k^c_{0l}+k^c_{0r}) (k^c_{rl}
   k^h_{lr}+k^c_{lr} k^h_{rl})]/(2\mathcal{N}),
\end{align}
and
\begin{align}
  D_{22}=&[-2 j_2^2
   (k^c_{l0}+k^c_{r0}+k^c_{0l}+k^c_{rl}+k^c_{0r}+k^c_{lr}+k^h_{rl}+k^h_{lr})+2 j_2 (k^c_{lr} k^h_{rl}-k^c_{rl} k^h_{lr})+k^c_{l0}
   k^c_{rl} k^c_{0r}\nonumber\\ &+k^c_{r0} k^c_{0l} k^c_{lr}+k^c_{0l} k^c_{rl}
   k^h_{lr}+k^c_{0l} k^c_{lr} k^h_{rl}+k^c_{rl} k^c_{0r}
   k^h_{lr}+k^c_{0r} k^c_{lr} k^h_{rl}]/(2\mathcal{N}),
\end{align}
with
\begin{align}
  \mathcal{N}\equiv (k^c_{l0}
   (k^c_{r0}+k^c_{rl}+k^c_{0r}+k^h_{rl})+k^c_{r0}
   (k^c_{0l}+k^c_{lr}+k^h_{lr})+(k^c_{0l}+k^c_{0r})
   (k^c_{rl}+k^c_{lr}+k^h_{rl}+k^h_{lr}).
\end{align}
The cumulants of heat and work then follow from the expressions
Eqs.~\eqref{eq:solarpower}-\eqref{eq:solardc} in the main text. 

The tightness of the resulting bounds on power \eqref{eq:b1}, \eqref{eq:b3} and
\eqref{eq:b4} are explored in Fig.~\ref{fig:solarcell2} as a function of the
two model parameters $\varGamma_c$ and $\varGamma_h$. Each of the three bounds can
become the tightest for some combination of parameters, showing that the
bounds are independent for this model. The ratio to the actual power becomes
small for small $\varGamma_c$, when nonradiative transitions between the two
levels are suppressed, such that a tight coupling between heat and work is
established.

\section{Brownian gyrator}
\label{sec:gyrator}

\begin{figure}
  \centering
  {\includegraphics[width=0.5\textwidth]{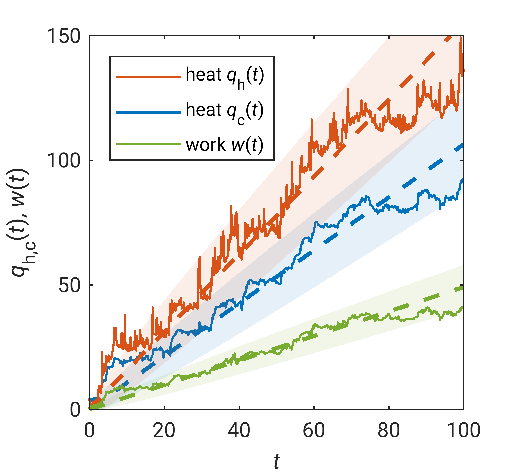}}
    \caption{Trajectories of the
    integrated heat $q_{h,c}(t)$ and work $w(t)$. The mean values and
    standard deviations according to the analytical solution are shown as
    dashed lines and shaded regions, respectively. Parameters are $T_c=1$,
    $T_h=7$, $\mu=1$, $u_1=1$, $u_2=1.2$, $k=k_s/2$, $c=0.8$. }
  \label{fig:gyrator2}
\end{figure}

The Langevin
equation for the Brownian gyrator, \eqref{eq:langevin_gyrator} in the main text, can be written as
\begin{equation}
  d\vec{x}(t)=-\vec{Ax}dt+\sqrt{2\vec{D}}d\vec{W}(t),
\end{equation}
with the non-symmetric matrix
\begin{equation}
   \vec{A}\equiv\left(
  \begin{matrix}
    a_{11}&a_{12}\\a_{21} &a_{22}
  \end{matrix}\right)
\equiv\mu\left(
  \begin{matrix}
    u_1 &c-k\\c+k &u_2
  \end{matrix}\right)
\label{eq:Amatrix}
\end{equation}
subsuming the potential force and the non-conservative, external force, the Einstein relation
$\vec{D}\equiv \mu \diag(T_c,T_h)$, and a two-dimensional standard Wiener process
$\vec{W}(t)$. In the following, we re-scale time such that $\mu=1$.
The differentials for work and heat, Eqs.~\eqref{eq:dw}-\eqref{eq:dqh}, are
identified as
\begin{equation}
  \label{eq:work}
  \dbar w(t)=d\vec{x}^Tk\vec{S}\vec{x}=-\vec{x}^T\vec{A}^Tk\vec{S}\vec{x}\,dt-\vec{x}^Tk\vec{S}\sqrt{2\vec{D}}d\vec{W},
\end{equation}
with the matrix
\begin{equation}
   \vec{S}\equiv\left(
  \begin{matrix}
    0&-1\\1 &0
  \end{matrix}\right)
\end{equation}
and 
\begin{equation}
  \dbar q_{c,h}(t)=d\vec{x}^T\vec{Q}(-\vec{A}\vec{x})=\vec{x}^T\vec{A}^T\vec{Q}\vec{A}\vec{x}\,dt-\vec{x}^T\vec{A}^T\vec{Q}\sqrt{2\vec{D}}d\vec{W},
\end{equation}
where we choose $\vec{Q}=\diag(1,0)$ for the heat $q_c$ dissipated into the
cold reservoir and $\vec{Q}=\diag(0,-1)$ for the heat $q_h$ extracted from the
hot reservoir. We employ the Stratonovich convention for stochastic integrals
throughout, e.g., for the calculation of the integrated work and heat from
above differentials. Sample trajectories for the integrated heat and work are
shown in Fig.~\ref{fig:gyrator2}.

Work and heat are connected through the first law
\begin{equation}
  \label{eq:1stlaw2}
  \dbar q_h-\dbar q_c-\dbar w=d[V(\vec x)],
\end{equation}
moreover, we find the relation
\begin{equation}
  \label{eq:heatrel}
  a_{12}\dbar q_h-a_{21}\dbar q_c=\frac{1}{2}d[a_{11}a_{21}x_1^2+a_{22}a_{12}x_2^2+2a_{12}a_{21}x_1x_2].
\end{equation}
Since the right hand sides of Eqs.~\eqref{eq:1stlaw2} and \eqref{eq:heatrel}
are total differentials of functions that do not scale with time, we conclude
that the fluctuations of the heat currents and the work are mutually dependent
on large time scales. It is therefore possible to focus on the current
$q_c(t)$ only (setting $\vec{Q}=\diag(1,0)$ in the following), with mean $j_c$
and diffusion coefficient $D_c$. The mean and diffusion of the current
$q_h(t)$ then follow readily as $j_h=(a_{21}/a_{12})j_c$ and
$D_h=(a_{21}/a_{12})^2D_c$, respectively, and likewise for the work
$j_w=(a_{21}/a_{12}-1)j_c$ and $D_w=(a_{12}/a_{21}-1)^2D_c$. These
considerations yield the efficiency
\begin{equation}
  \eta=(a_{21}-a_{12})/a_{21}=2 k/(c+k)
  \label{eq:effsupp}
\end{equation}
(Eq.~\eqref{eq:gyratoreff} in the main text) and the equality
\begin{equation}
  \frac{D_w}{j_w^2}=\frac{D_c}{j_c^2}=  \frac{D_h}{j_h^2},
\end{equation}
c.f.~Eq.\eqref{eq:unc_gyr}.

The fluctuations of the heat $q_c(t)$ can be derived from the scaled cumulant
generating function
\begin{equation}
  \label{eq:scgf2}
  \lambda(z)=\lim_{t\to\infty}\frac{1}{t}\ln\mean{\exp(zq_c(t))}=j_cz+D_c z^2+\mathcal{O}(z^3)
\end{equation}
with a real parameter $z$.
Employing standard methods, as recently pedagogically reviewed in
\cite{touc17}, $\lambda(z)$ follows as the largest eigenvalue of the tilted
adjoint Fokker-Planck operator
\begin{equation}
  \label{eq:FP}
  \mathcal{L}^\dagger(z)\equiv -\vec{x}^T\vec{A}^T\vec{\nabla}+\vec{\nabla}^T\vec{D}\vec{\nabla}+z\vec{x}^T\vec{A}^T\vec{Q}\vec{A}\vec{x}-z\vec{\nabla}^T\vec{D}\vec{Q}\vec{A}\vec{x}-z\vec{x}^T\vec{A}^T\vec{Q}\vec{D}\vec{\nabla}+z^2\vec{x}^T\vec{A}^T\vec{Q}\vec{D}\vec{Q}\vec{A}\vec{x}
\end{equation}
with $\vec{\nabla}\equiv\partial/\partial\vec{x}$.
For $z=0$, the largest eigenvalue is $\lambda(0)=0$ with the corresponding
eigenfunction $g(\vec{x},z=0)=1$. For general $z$, we use the Gaussian ansatz
$g(\vec{x},z)=\exp[-\vec{x}^T\vec{G}(z)\vec{x}/2]$ with a symmetric matrix $\vec{G}(z)$ for the eigenfunction, leading to
\begin{align}
\lambda(z)=(\mathcal{L}^\dagger(z)g(\vec{x},z))/g(\vec{x},z)=&\vec{x}^T\vec{A}^T\vec{G}\vec{x}+\vec{x}^T\vec{G}\vec{D}\vec{G}\vec{x}-\tr(\vec{D}\vec{G})+z\vec{x}^T\vec{A}^T\vec{Q}\vec{A}\vec{x}\nonumber\\
&+z\vec{x}^T\vec{G}\vec{D}\vec{Q}\vec{A}\vec{x}-z\tr(\vec{D}\vec{Q}\vec{A})+z\vec{x}^T\vec{A}^T\vec{Q}\vec{D}\vec{G}\vec{x}+z^2
 \vec{x}^T\vec{A}^T\vec{Q}\vec{D}\vec{Q}\vec{A}\vec{x}
\end{align}
where we suppress the argument of $\vec{G}(z)$ here and in the following for readability. 
Comparing the coefficients of the quadratic form then yields
\begin{equation}
  \lambda(z)=-\tr(\vec{D}\vec{G})-z\tr(\vec{D}\vec{Q}\vec{A})
  \label{eq:lambda}
\end{equation}
and
\begin{equation}
  (\vec{G}\vec{A})_s+\vec{G}\vec{D}\vec{G}+z\vec{A}^T\vec{Q}\vec{A}+2z(\vec{G}\vec{D}\vec{Q}\vec{A})_s+z^2\vec{A}^T\vec{Q}\vec{D}\vec{Q}\vec{A}=0,
  \label{eq:G}
\end{equation}
where we denote the symmetrization of a matrix $\vec{X}$ as
$(\vec{X})_s\equiv(\vec{X}+\vec{X}^T)/2$. Eq.~\eqref{eq:G} presents three conditions for the
three independent coefficients of $\vec{G}$. The Gaussian ansatz thus yields indeed
an eigenfunction of $\mathcal{L}^\dagger(z)$. However, since Eq.~\eqref{eq:G}
is nonlinear in $\vec{G}$, it is hard to find a general solution for $\vec{G}$ for
arbitrary $z$. Nonetheless, Eqs.~\eqref{eq:G} and \eqref{eq:lambda} can be
reduced to a set of linear equations by expanding $\lambda(z)$ and $\vec{G}(z)$ as
Taylor series \eqref{eq:scgf2} and
\begin{equation}
  \vec{G}(z)=z\vec{G}_1+z^2\vec{G}_2+\mathcal{O}(z^3),
\end{equation}
respectively. The first order in $z$ yields
\begin{equation}
  j_c=-\tr(\vec{D}\vec{G}_1)-\tr(\vec{D}\vec{Q}\vec{A})
\end{equation}
with $\vec{G}_1$ determined by
\begin{equation}
  (\vec{G}_1\vec{A})_s+\vec{A}^T\vec{Q}\vec{A}=0
\end{equation}
and the second order in $z$ yields
\begin{equation}
  D_c=-\tr(\vec{D}\vec{G}_2)
\end{equation}
with $\vec{G}_2$ determined by
\begin{equation}
  \vec{G}_1 \vec{D} \vec{G}_1+2(\vec{G}_1\vec{D}\vec{Q}\vec{A})_s+\vec{A}^T\vec{Q}\vec{D}\vec{Q}\vec{A}+(\vec{G}_2\vec{A})_s=0.
\end{equation}
Solving these linear systems of equations, we finally obtain
\begin{equation}
  j_c=\frac{a_{12}}{\tr \vec{A}}(a_{12}T_h-a_{21}T_c)
\end{equation}
and
\begin{equation}
  D_c=\frac{a_{12}^2}{(\tr
    \vec{A})^3}[a_{21}^2T_c^2+T_cT_h(a_{11}^2+a_{22}^2+2\det \vec{A})+a_{12}^2T_h^2].
\end{equation}

\begin{figure}
  \centering
  {\includegraphics[width=0.9\textwidth]{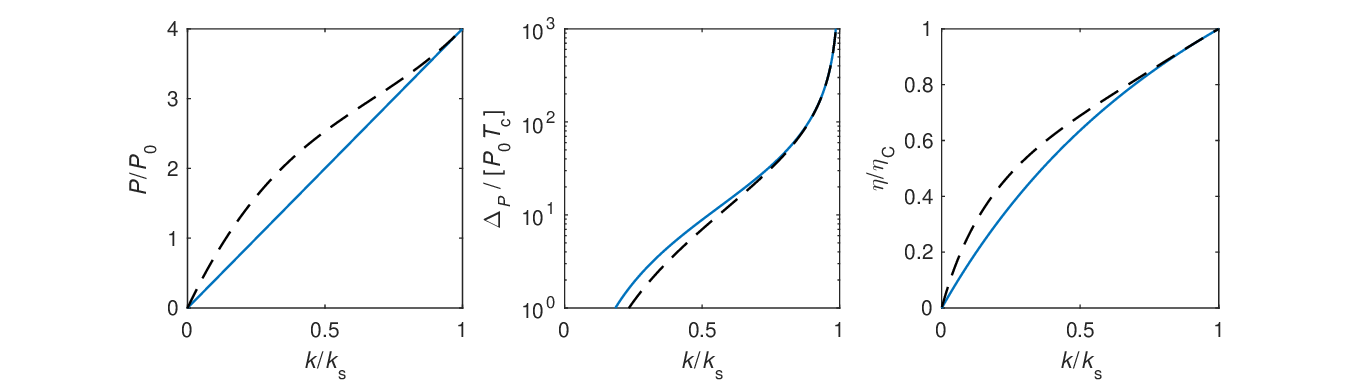}}
    \caption{Analytical results for the power (left), constancy (middle) and
      efficiency (right) of the Brownian gyrator as a function of the
      load parameter $k$. Unlike in
      Fig.~\ref{fig:gyrator}b of the main text, we here scale the
      mobility according to $\mu=\mu_0/(1-k/k_s)$, which leads to
      finite power at Carnot efficiency at the cost of diverging power
      fluctuations. The actual values for power, constancy and
      efficiency are shown in blue, the black dashed curves are the
      bounds derived from Eq.~\eqref{eq:b1} in the main text and the
      respective other two quantities. Parameters: $T_h/T_c=7$, $u_1/c=1$, $u_2/c=1.2$, $P_0\equiv \mu_0 k_s^2(T_c+T_h)/(2(u_1+u_2))$.}
  \label{fig:gyrator3}
\end{figure}

Restoring the original units and notation \eqref{eq:Amatrix}, we obtain 
the expressions \eqref{eq:gyrpower} for the power and the explicit result
Eq.~\eqref{eq:unc_gyr} for the uncertainty relation. In particular,
for the constancy of power we obtain
\begin{equation}
  \varDelta_P=2D_w=\frac{8\mu k^2
    T_c^2}{(1-\eta_C)(u_1+u_2)}\left[1+\frac{[c\eta_C(1-k/k_s)]^2}{(u_1+u_2)^2(1-\eta_C)}\right]
\label{eq:gyratorconstancy}
\end{equation}
with the stall-parameter $k_s=c\eta_C/(2-\eta_C)$. The
efficiency~\eqref{eq:effsupp} reaches the Carnot limit for $k\to
k_s$. With all other parameters kept fixed, the constancy remains
finite in this limit, such that the bound~\eqref{eq:b1} demands
vanishing power as visible in Fig.~\ref{fig:gyrator}b of the main
text. Yet, finite power can be achieved in this limit if the mobility
$\mu$ is scaled up, which comes at the cost of diverging power
fluctuations~\eqref{eq:gyratorconstancy}. This behavior, which is
akin to what we have explored for the quantum dot solar cell in the
main text, is illustrated in Fig.~\ref{fig:gyrator3}.

Our bounds on efficiency and power become the tighter, the closer
$D_w\sigma/j_w^2$ is to unity. An \textit{upper} bound on this ratio that is
specific for the present model of the Brownian gyrator and $0<k<k_s$
follows from
\begin{equation}
  \frac{D_w\sigma}{j_w^2}=
  1+\frac{[c\eta_C(1-k/k_s)]^2}{(u_1+u_2)^2(1-\eta_C)} <
  1+\left[\frac{\sqrt{u_1u_2}}{u_1+u_2}\right]^2\frac{\eta_C^2}{1-\eta_C}\leq 1+\frac{\eta_C^2}{4(1-\eta_C)}
\end{equation}
and is approached for $k\to 0$, coupling $c\to \sqrt{u_1u_2}$ (larger
coupling leads to an unstable potential), and $u_1=u_2$.

\end{document}